\documentclass[12pt]{article}

\newcommand{\Lya}{Lyman-$\alpha$~}

\usepackage{scicite}

\usepackage{times}

\usepackage{color}
\usepackage{graphicx}

\newenvironment{sciabstract}{%
\begin{quote} \bf} 
{\end{quote}}

\newcounter{lastnote}

\title{A Local Clue to the Reionization of the Universe \footnote{This is the author's version of the work. It is posted here by permission of the AAAS for personal use, not for redistribution. The definitive version was published in Science {VOL 346,(10 October 2014)}, doi:10.1126/science.1254214}}

\author
 {Sanchayeeta Borthakur,$^{1{\dag}}$ Timothy M. Heckman,$^1$ Claus Leitherer,$^2$ Roderik A. Overzier$^3$\\
\\
\normalsize{$^{1}$Center for Astrophysical Sciences, Department of Physics \& Astronomy, }\\
\normalsize{Johns Hopkins University, Baltimore, MD 21218, USA}\\
\normalsize{$^{2}$Space Telescope Science Institute, 3700 San Martin Drive, Baltimore, MD 21218, USA}\\
\normalsize{$^{3}$Observat$\acute{o}$rio Nacional, Ministry of Science, Technology, and Innovation, Rio de Janeiro, Brazil}\\
\\
\normalsize{$^{\dag}$To whom correspondence should be addressed; E-mail:  sanch@pha.jhu.edu.}
}

\date{}

\begin{document} 


\baselineskip24pt


\maketitle


\begin{sciabstract}

Identifying the population of galaxies that was responsible for the reionization of the universe is a long-standing quest in astronomy. We present a possible local analog that has an escape fraction of ionizing flux of 21\%. Our detection confirms the existence of gaps in the neutral gas enveloping the starburst region. The candidate contains a massive yet highly compact star-forming region. The gaps are most likely created by the unusually strong winds and intense ionizing radiation produced by this extreme object. Our study also validates the indirect technique of using the residual flux in saturated low-ionization interstellar absorption-lines for identifying such leaky galaxies. Since direct detection of ionizing flux is impossible at the epoch of reionization, this is a highly valuable technique for future studies.

\end{sciabstract}

\newpage


The reionization of the universe is a crucial event in cosmic history. Identifying the source(s) responsible for reionization will allow us to understand the underlying physics.
Star-forming galaxies are the most likely candidates, but what is their nature? By what process does Lyman continuum radiation, capable of ionizing hydrogen atoms, escape the region of dense and cold gas from which the required population of hot massive stars forms? 

Studies to date have had limited success at detecting such candidates.
For example, Iwata et al. \cite{iwata09} found continuum leaking out of 17 galaxies from their sample of 198 star-forming galaxies (Lyman Break galaxies (LBGs) and Ly-alpha emitters (LAEs)) whereas Vanzella et al. \cite{vanzella10} found only one such leaking galaxy out of 102 LBGs. 
Most of the high redshift studies have only resulted in a handful of detections with significant escape fractions \cite{nestor13, mostardi13}. In some ways, the low incidence of such galaxies is not surprising. Hot massive stars are located in gas-rich regions with column densities ranging from 10$^{21}$ to 10$^{24}$~cm$^{-2}$. This is 4 to 7 orders-of-magnitudes higher than the column density of HI required to produce unit optical depth at the Lyman edge (N$\rm = 1.4~\times~10^{17}~cm^{-2}$). Therefore, in order to have a significant fraction of the ionizing flux escape from the galaxy, extreme conditions may be required.

Identifying local examples of such leaky galaxies will allow us to study in detail the processes that lead to the escape of Lyman continuum radiation. These local examples can also be used to validate indirect signatures of escaping ionizing flux that are based on the properties of the spectrum long-ward of the \Lya line (see below). This is invaluable, because the spectral region at shorter wavelengths is completely opaque at the epoch of recombination. 

We have previously uncovered a sample of relatively rare local (z$\sim$0.2) starburst galaxies whose overall properties such as mass, metallicity, size, morphology, star formation rate (SFR), kinematics, dust attenuation etc. are very similar to those of the LBGs  \cite{hoopes07, overzier10, overzier11, goncalves10, heckman11, basu-zych13}.  These galaxies are termed Lyman Break Analog galaxies (LBAs).
SDSS~J092159.38+450912.3 (hereafter J0921+4509) is one such local LBA ($z= 0.23499$) that is producing stars at the rate of 50 ~M$_{\odot}$~yr$^{-1}$. 
It belongs to a special class of LBA where several billion solar masses of stars are produced in an extremely compact central region of a radius of $\rm \sim100$~pc \cite{overzier09}. Such massive and highly compact star forming regions are referred to as dominant central objects (DCOs).
Due to their compact nature, DCOs can generate extreme winds that were found to propel the surrounding interstellar neutral and ionized gas outward at velocities in excess of 1000~km~s$^{-1}$ \cite[hereafter H11]{heckman11}.

Three independent but indirect lines of evidence were put forth by H11 \cite{heckman11} suggesting that a significant amount of ionizing radiation could be escaping from this galaxy and others with DCOs. First, these authors found that the strongest interstellar absorption-lines tracing the neutral phase of the interstellar medium (ISM) were optically thick (saturated) but still had a significant residual flux in the line cores ($\sim$20-30\%; see Figure~4 and 5 in H11 \cite{heckman11}). This is an indication that the far-UV continuum source, the DCO, is only partially covered by optically thick neutral material. Second, they detected a significant amount of blue-shifted \Lya emission, in spite of the presence of neutral gas seen in absorption at those velocities. This requires only partial coverage of the starburst region by the outflowing neutral gas. Third, the star formation rate estimated for this galaxy based on the extinction-corrected H$\alpha$ emission line was several times smaller than the value based on the sum of the far-IR and far-UV continuum luminosities. This could indicate that some of the ionizing flux escapes the galaxy and thus does not produce H$\alpha$ emission by photoionization and recombination. These properties led H11 \cite{heckman11} to speculate that the extreme feedback provided by an extraordinarily compact massive starburst (like a DCO) might be required to enable the escape of ionizing radiation from galaxies.

In order to directly confirm the inference of escaping ionizing radiation and compare it to that predicted by the residual flux technique, we conducted a program with the Cosmic Origins Spectrograph (COS) aboard the Hubble Space Telescope (HST) to obtain the spectrum of J0921+4509 below the Lyman edge at 912~$\rm \AA$ in the rest-frame of the galaxy.
The data were obtained in two separate sets of observations of duration 4 and 5 orbits of HST respectively, which were reduced using the standard COS data reduction pipeline \cite{cosihb}.
The final spectrum (Fig.~1) was produced by binning 150 pixels of the weighted average of the two data sets in order to improve on the signal-to-noise. 
The spectrum was corrected for the Milky Way's dust extinction in the direction of J0921+4509.  
 We report a flux density detection of  $\rm 3.7 (\pm 0.8) \times 10^{-17} ~ erg~s^{-1}~cm^{-2}~\AA^{-1}$ just below the Lyman edge. This corresponds to a luminosity, $\rm \lambda~L_{\lambda}$, at rest wavelength of $\rm \lambda=910~\AA$ of $\rm 5.0 \times 10^{42}$~erg~s$^{-1}$.

The young stars that produce the ionizing flux also produce strong stellar winds, which can be traced using transitions such as N~V and C~IV. These transitions are minimally contaminated by interstellar features and have the greatest diagnostic power to identify the underlying stellar populations.
The strong P-Cygni profiles in these transitions (Fig.~2) clearly indicate the presence of young massive O stars capable of producing substantial amounts of ionizing radiation.
A stellar population model \cite{starburst99} of an instantaneous burst of 3~Myrs old and Kroupa initial mass function describes our data best, thus, suggesting that J0921+4509 has a large fraction of extremely young and massive stars.

Before describing how we quantified the escaping ionizing radiation in this galaxy, it is important to distinguish between two sources of opacity provided by the galaxy's ISM. The first is the dust. Opacity due to dust suppresses the amount of escaping ultraviolet light by an amount that gradually and smoothly increases with decreasing wavelength. It therefore renders the ultraviolet continuum fainter and redder, but does not produce any signature across the Lyman edge at 912~$\rm \AA $. The second
source of opacity is associated with photoelectric absorption by neutral Hydrogen, which will produce a sharp drop in flux density at the Lyman edge. For simplicity,  we will describe these two sources of opacity separately. 

\begin{figure}[!h]
\center
\includegraphics[trim=0mm 0mm 0mm 0mm, clip=true, scale=0.7]{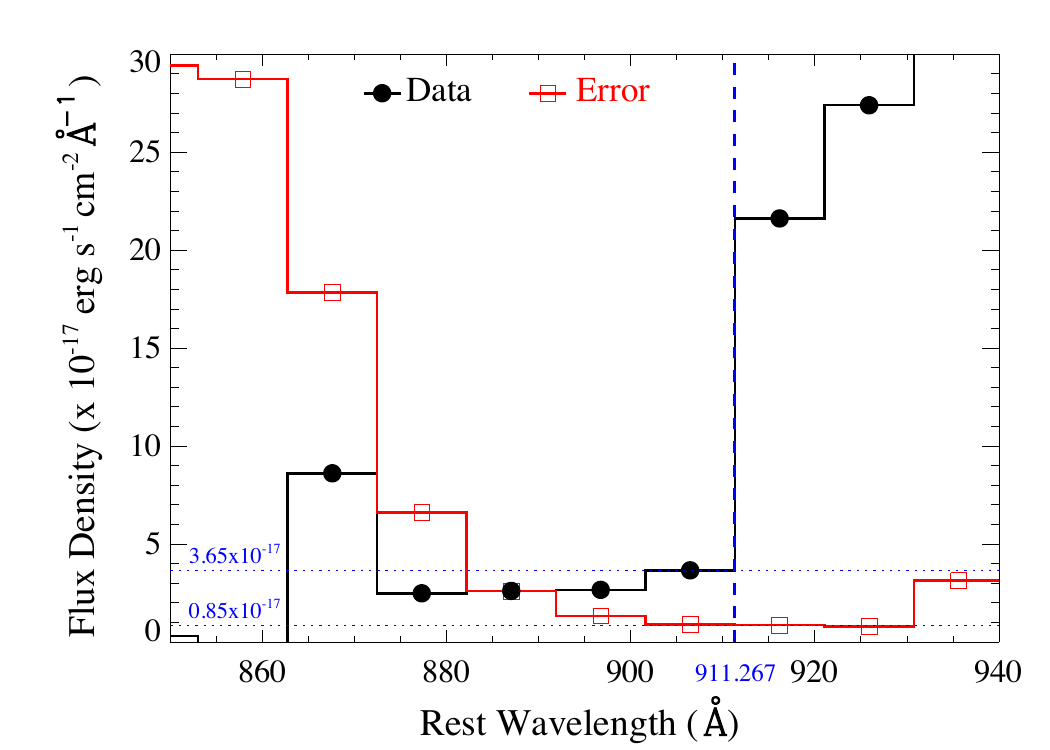}
\caption{COS spectra of J0921+4509 showing flux density below rest-frame 912~$\rm \AA$ in black (filled circles) and the associated errors in red (open squares). The flux density and errors were obtained by binning the  COS pipeline data by 150 pixels $\sim \rm 10~\AA$ to increase the signal-to-noise ratio.
The dashed blue line marks the position of the Lyman continuum break ($\rm \equiv 13.6~eV$). 
We detected Lyman continuum flux between wavelengths 890-910~$\rm \AA$. The flux at 910 $\rm \AA$ was measured to be  $\rm 3.7 \pm 0.8 \times 10^{-17}~erg~s^{-1}cm^{-2}\AA^{-1}$. \label{lyc_flux}  }
\end{figure}

First, we consider the effects of photoelectric absorption in the absence of dust (the dust-free case).  The stellar population model predicts the intrinsic ratio of ionizing to non-ionizing flux, i.e.,,  the drop in flux at the Lyman edge in the absence of photoelectric absorption by the interstellar medium. 
A comparison of the drop in flux density between our data and the best stellar population model describing our data (Fig.~\ref{data_vs_models}) gives us a measure of the fraction of Lyman continuum photons that suffer photoelectric absorption by atomic hydrogen in the ISM of the galaxy. In mathematical terms, 
\begin{equation}\label{f_dustfree}
f_{esc}^{dust free} = \Big( \frac{F_{\lambda912-}}{F_{\lambda912+}} \Big)_{obs}   \Big/ ~\Big( \frac{F_{\lambda912-}}{F_{\lambda912+}} \Big)_{norm.model}
\end{equation}
where, $ {F_{\lambda912-}}$ and ${F_{\lambda912+}}$ are the fluxes below and above 912~$\rm \AA$. ${F_{\lambda912-}}$ was measured by averaging the flux density over 10$\rm \AA$ below the Lyman break (see Fig.\ref{lyc_flux}). To avoid the effects of the confluence of the stellar and interstellar Lyman series absorption-lines near the Lyman edge, we measured $F_{\lambda912+}$ in a window between 1000-1020 \AA\ (beyond Lyman~$\gamma$).
The observed drop in flux density compared to the stellar population model implies a $f_{esc}^{dust free} \equiv  21\pm 5 \%$.
The inferred incompleteness in coverage of the starburst region by optically-thick neutral material predicts $f_{esc}^{dust free}\equiv 1-f_{cov}$, which in turn is equal to the residual flux.  Our measurement of  $f_{esc}^{dust free}$ is in good agreement with the estimate based on the residual flux ($\sim$20-30\%) in the strongest interstellar absorption-lines \cite[H11]{heckman11}.

We can now consider the effect of dust absorption. Here we are concerned with estimating what we call the absolute escape fraction of Lyman continuum photons, due to the combined effect of photoelectric absorption and dust extinction. Thus, we want to estimate the ratio between the observed luminosity of the escaping ionizing radiation to the intrinsic luminosity, in the absence of both dust and neutral hydrogen. 
The amount of dust absorption in the far-ultraviolet is substantial in this galaxy: the ratio of ultraviolet to infrared flux is just $\sim$ 10\% \cite{overzier11}. Assuming that the infrared (IR) and ultraviolet (UV) luminosities dominate the bolometric luminosity ($\rm L_{bol} \sim L_{UV}+L_{IR}$), we adopt the definition of absolute escape fraction, $f_{esc, abs}$ as described in \cite{grimes09}
\begin{equation}\label{eq_f_abs}
f_{esc, abs} = \rm \frac{L_{ioniz,esc}}{L_{UV}+L_{IR}} ~  \Big( \frac{L_{bol}}{L_{ioniz}} \Big)_{intrinsic}
\end{equation}
We estimate the escaping Lyman continuum luminosity, $\rm L_{ioniz,esc}$, from the data by measuring the parameter $\rm \lambda~L_{\lambda}$ just below the Lyman continuum break and correcting it by a factor as follows
\begin{equation}\label{eq_Lion_esc}
\rm L_{ioniz,esc}=  \Big({\lambda~L_{\lambda}} \Big)_{\lambda=912^-\AA}^{observed}~~ \Bigg( \frac{L_{ioniz}}{(\lambda~L_{\lambda})_{\lambda=912^-\AA}} \Bigg)^{intrinsic}
\end{equation}
The best-fit stellar population model predicts the intrinsic ratio $\rm L_{bol}/L_{ioniz}$ to be around 4, and the correction factor of Eq.~\ref{eq_Lion_esc} to be 0.9. This leads to an estimate of $f_{esc, abs} \sim 1\%$.

The difference between $ f_{esc}^{dust free} $ and $ f_{esc, abs}$ is due to the absorption of ionizing radiation by dust. It also indicates that the covering fractions of the neutral gas and the dust are not the same. While the distribution of neutral gas evidently has gaps/holes, the dust almost fully covers the entire star-forming region. Based on this observation, we conclude that most of the dust absorption in this galaxy is associated with the ionized gas, consistent with the arguments given in H11 \cite{heckman11}. Thus, strong feedback from the starburst, in the form of ionizing radiation and shock heating by the galactic wind, is able to fully ionize holes in the distribution of neutral gas that envelops the starburst region. In these holes, the gas is fully ionized and allows the Lyman continuum photons to escape. However, the dust survives and is able to absorb both the ionizing and non-ionizing ultraviolet photons (as shown in the schematic diagram in Fig.~\ref{diagram}). 
The holes contain ionized gas with negligible optical depth due to photoelectric absorption for Lyman continuum photons, and also dust.

How relevant is the knowledge gained from J0921+4501 for understanding the sources responsible for reionization, given that it is highly obscured by dust?
The good news is that the physics behind the creation of the holes that allow Lyman continuum to escape may be valid for any compact starburst.
Such compact starburst galaxies are commonly seen at high redshifts \cite{huang13}. 
 In fact, the highest redshift galaxy (z=10.7) known so far has a size ($\sim$200~pc) and stellar mass ($10^{8-9}~M_{\odot}$) \cite{coe12} similar to the starburst in J0921+4501.
On the other hand, mounting evidence from deep surveys of the early universe with the HST imply that the amount of dust absorption in star-forming galaxies is very small at the highest redshifts \cite{bouwens13}. This is most plausibly caused by the much lower metallicities (and hence lower dust-to-gas ratios) in these early galaxies. 
Recently, an intensely star-forming galaxy with very low dust and metals at redshift $z\sim 7$ was discovered \cite{ouchi13}.
Therefore, while dusty objects just like J$0921+4501$ would not be good candidates for the population of galaxies that reionized the universe, similar analogs from the early universe that are compact but nearly dust-free would be excellent candidates.

It is important to reemphasize that the opacity due to neutral hydrogen in the intergalactic medium (IGM) during the epoch of reionization ($z\sim 6-11$) is so large that there is no hope of directly observing flux below the Lyman edge from these early galaxies and directly measuring the escape fraction. We have briefly described above the interesting idea that the amount of residual flux in the cores of saturated interstellar absorption lines that trace the neutral gas can serve as a proxy measurement of the escape fraction. This was proposed by Heckman et al. in 2001 \cite{heckman01}, then applied by Grimes et al. in 2009 \cite{grimes09} to a sample of local starbursts, and then applied by H11 \cite{heckman11} to a sample of LBAs. This idea has recently been extended to high-redshift galaxies by  Jones et al. \cite{jones13}. In this work, we have validated this indirect technique.  
The direct value we measure for $ f_{esc}^{dust free}$ in J$0921+4509$ is similar to the indirect estimate in H11 \cite{heckman11}.

\bibliographystyle{science}	        

\section*{Acknowledgement}
We thank the referees for their insightful suggestions.
We thank A. Aloisi for assistance during the observations of this project. 
S.B is supported by a grant under program number 12886 provided by NASA through a
grant from the Space Telescope Science Institute, which is operated by the 
Association of Universities for Research in Astronomy, Inc., under NASA
contract NAS5-26555. Data used for this paper is publicly available in the 
Mikulski Archive for Space Telescopes (MAST) under program 12886 and 11727.\\

\clearpage

  Properties of The Target Galaxy \cite{overzier11}.\\
\begin{tabular}{c c c c c c c  c c r }

\hline
Galaxy  & Redshift & Stellar Mass & DCO Mass    & Metallicity      & SFR                                          & Dust   Content \\ 
               &                 &  Log~M$_*$ &  Log~M$_*$ & 12+log(O/H) & H$\rm \alpha+24\mu m $          &  $\beta_{rest}$  \\
               &                 &  (M$_\odot$ )&     (M$_\odot$ )              &      &        (M$\rm _\odot~yr^{-1}$) &                             \\ 
\hline
\hline
J0921+4509 &  0.235  &  10.8  &  9.0$-$9.2  & 8.67    &  55.1  &    -1.37    \\ 
\hline
\end{tabular}

\clearpage

\begin{figure}[!h]
\hspace{-0.5cm}
\includegraphics[trim=5mm 00mm 0mm 0mm, clip=true, scale=0.7]{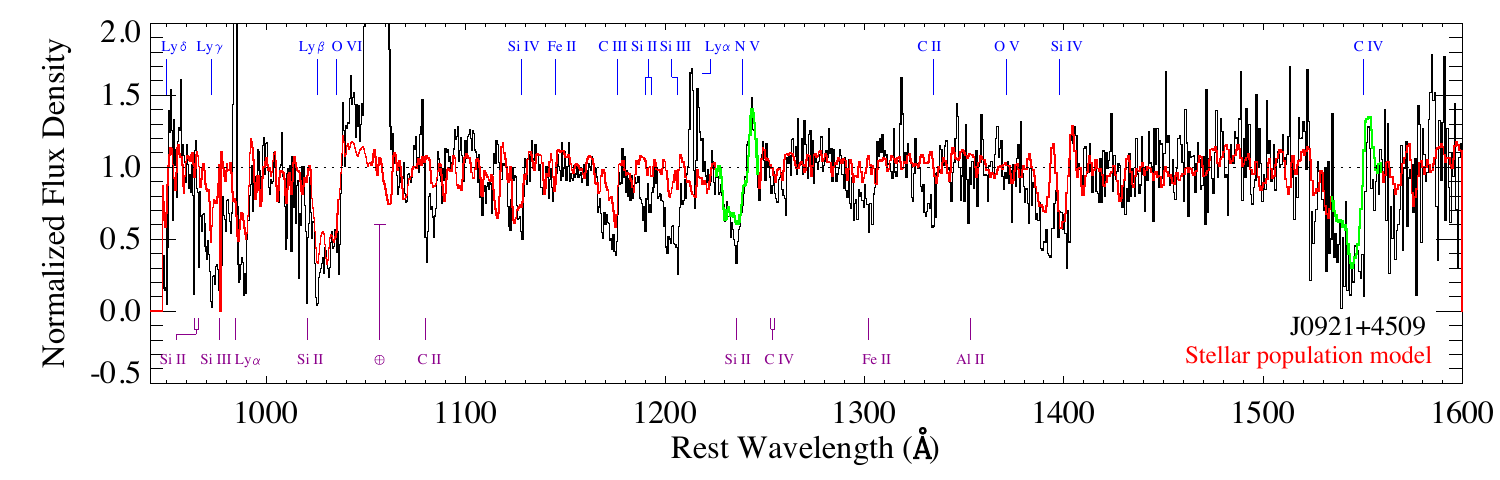}
\caption{Flux density normalized combined spectrum from COS G140L (1040-1600~$\rm \AA$) and G130M ($\sim$950-1040~$\rm \AA$) of LBA J0921+4509 (black) overlaid with the best fit stellar population model (Starburst99) of instantaneous star formation of 3~Myr, solar metallicity, and Kroupa initial mass function (red). We fitted the high-ionization stellar wind transitions (shown in green) to evaluate the best stellar population model. The transitions associated with J0921+45 are marked in blue above the spectrum, and Milky Way \&  geocoronal features are marked in purple in the space below the spectrum. The J0921+45 spectrum shows prominent P-Cygni profiles from the stellar winds (e.g., O~VI$\lambda\lambda$1032,1038, N~V$\lambda\lambda$1239,1243, C~IV$\lambda\lambda$1548,1551). This confirms the presence of a young stellar population with massive O stars in this galaxy.  \label{population}  }
\end{figure}

\begin{figure}[!h]
\center
\includegraphics[trim=0mm 00mm 0mm 0mm, clip=true, scale=0.7]{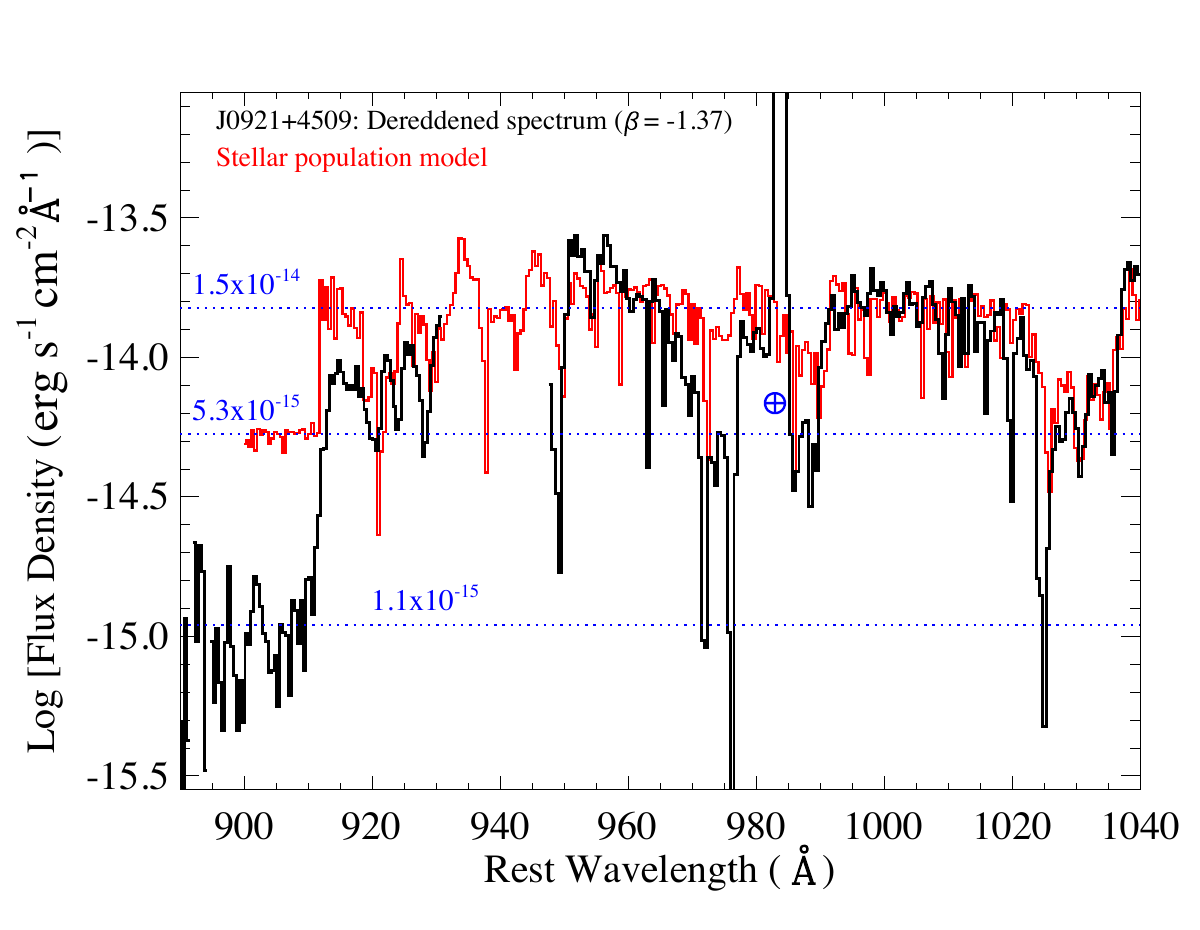}
\caption{The COS spectrum of J0921+4509 between 890-1040~\AA\ in the rest-frame of the galaxy shown in black. The spectrum was corrected for dust attenuation \cite{leitherer02} corresponding to the observed UV slope of $\beta=-1.37$ in J0921+4509. The fit-best stellar population model is shown in red and has been normalized to match flux density of the data between 960-1040~$\rm \AA$. Dotted lines indicate the observed and predicted (by model) flux densities at the Lyman edge (900 to 910 ) and at wavelengths between 1000 to 1020 ~$\rm \AA$.  The ratio of strength of the observed drop to the theoretical drop represents the dust free escape fraction.}  \label{data_vs_models}  
\end{figure}

\begin{figure}[!h]
\begin{center}
\includegraphics[trim=0mm 00mm 0mm 0mm, clip=true, scale=0.6]{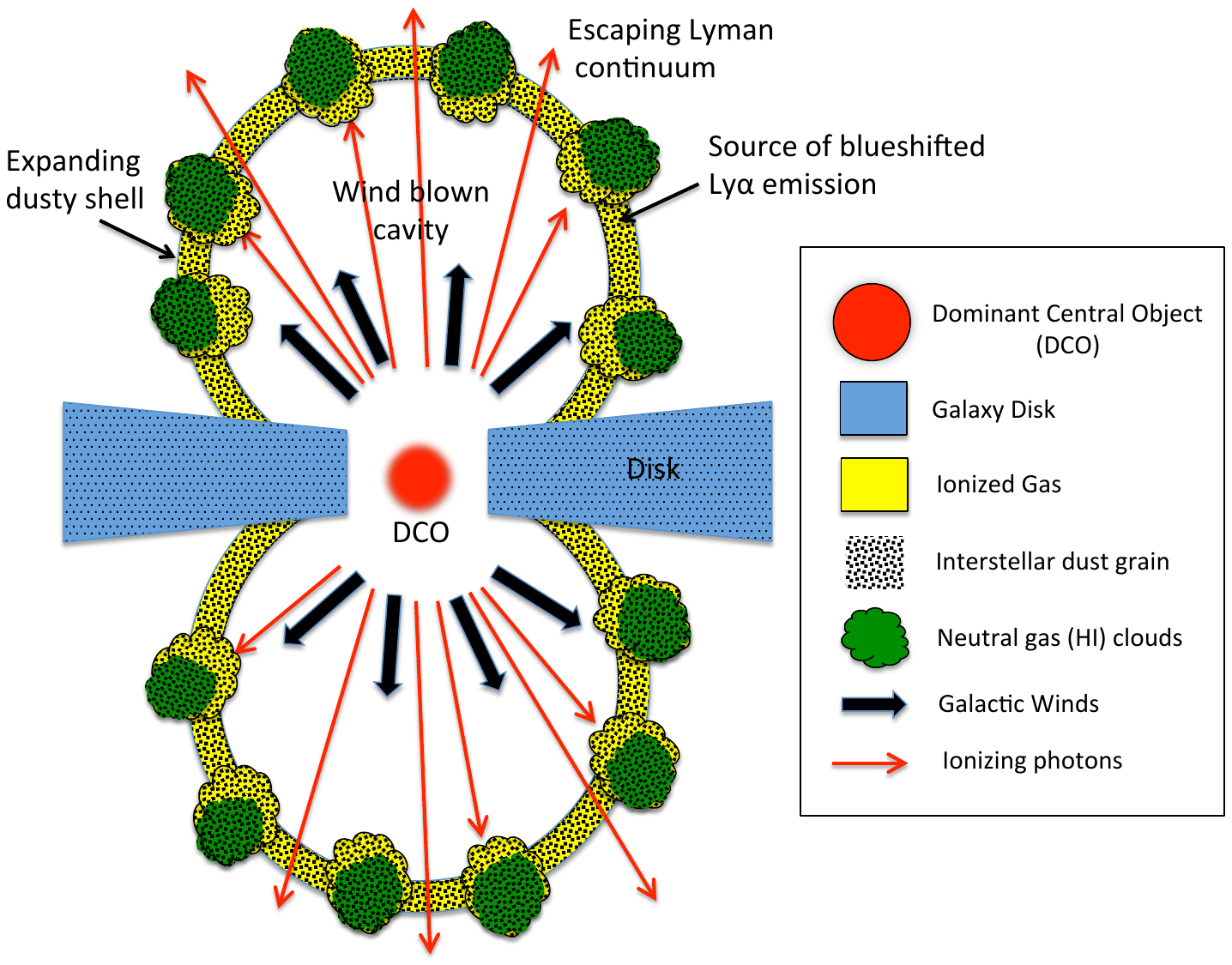}
\end{center}
\caption{Schematic diagram of the star-forming region associated with a DCO such as in J0921+4509. The extreme conditions in the DCO generate a strong galactic wind and large quantities of ionizing flux that clear out a cavity (wind-blown cavity in the figure) and parts of the neutral gas cover. However, the  entire region is covered by dust. The neutral gas cover contains gaps, where the gas gets ionized completely and consequently has negligible optical depth due to photoelectric absorption of Lyman continuum photons by hydrogen.  Thus, these gaps facilitate the escape of Lyman continuum and \Lya photons.  
 \label{diagram}  }
\end{figure}

\end{document}